# A Generalized Short Circuit Ratio for Multi-Infeed LCC-HVDC System

Feng Zhang, Huanhai Xin, Zhen Wang, Deqiang Gan, Qian Xu, Pan Dai, Feng Liu

***Abstract*—The relationship between the short circuit ratio (SCR) and static voltage stability is analyzed in this paper. According to eigenvalue decomposition method, a novel concept named generalized short circuit ratio (gSCR) has been proposed for multi-infeed LCC-HVDC (MIDC) systems to mathematically measure the connected AC strength from the point view of voltage stability, which can overcome the rule-of-thumb basis of existing multi-infeed short circuit ratio (MISCR) concept. In gSCR, two indices, the critical gSCR (CgSCR) and the boundary gSCR (BgSCR) are developed to quantitatively evaluate if the connected AC system is strong or weak, in which CgSCR=2 and BgSCR=3 are two critical values for strength evaluation. Finally, numerical simulations are conducted to validate the effectiveness of the proposed gSCR concept.**

## I. INTRODUCTION

Recently, the HVDC transmission is increasingly used as the power transmission way for widespread renewable energy generation [1]. In China, several ±800kV HVDC transmission projects have been particularly put into service for long-distance and sending-out electric power from large hydro-power plants and high-density wind farms to form the large North-, East- and Central China grid jointly with other ultra-high-voltage (UHV) AC transmission, while the total installed capacity will exceed 700GW by 2020 [2]. These close HVDCs constitute the multi-infeed HVDC system (MIDC) with AC power grid [3]. During the past decades, concerns on MIDC were mainly focused on the power transmission limitation and the voltage stability issues, e.g., stability mechanism, strategies to improve dynamic performance, etc, which have been carried over from similar historical concerns on single-infeed HVDC (SIDC) system [4].

In theoretical research and engineering application, the short circuit ratio (SCR) has been proposed to measure the strength and stability of the connected AC system [4]. The inherent issue of HVDC converters is recognized as voltage instability in MIDCs, similar to that of the SIDC system. Many HVDC operation problems, e.g, concurrent commutation failure of MIDC converters in close electrical proximity, transient overvoltage (TOV) at the converter ac buses and harmonic instability were believed to be affected by voltage instability [5]-[7]. A SCR index was developed to evaluate the strength of connected AC system for single-indeed HVDC system. To consider the MIDC emerged later, a multi-infeed short circuit ratio (MISCR), as an extension of SCR was further proposed by *CIGRÉ* to give a quantitative assessment of the strength of AC/DC system considering the neighboring HVDC's voltage influence [5]. However, MISCR is a rule-of-thumb extension of SCR concept and it is lacking of strict theory basis [7].

Several attempts have been made to give theoretical insight into physical mechanism of voltage stability, power system planning and terminal locations selection of MIDC [7-13]. An analytic equivalent of the empirical index in [7] is derived to facilitate rigorous analysis of voltage/power interactions in MIDC systems, which is proposed to give theoretical explanation of the voltage stability. In [8], power system planning and location selection is solved by analyzing an optimal problems of dc-segmentation for MIDC systems based on stability performance, and so on. But these works haven't essentially explain the physical mechanism of MISCR and the relationship between the MISCR and static voltage stability.

In this paper, a novel generalized short circuit ratio is proposed, which is derived by the static voltage stability analysis and the eigenvalue decomposition. Firstly, the relationship between SCR and static voltage stability is analyzed. The characteristic equation is then yielded by analogy analyzing between SIDC and MIDC. The gSCR concept is further developed to mathematically measure the connected AC strength from the point view of voltage stability. Finally, the effectiveness of the proposed index is verified by simulations on the platforms of Matlab and DIgSILENT, respectively.

## II. RELATIONSHIP BETWEEN SCR AND STATIC VOLTAGE STABILITY

### A. *The definition and characteristics of SCR*

In a SIDC system, the definition of SCR is

$$SCR = S_{ac}/P_{dN} = U_N{}^2/(P_{dN}\cdot Z) = 1/(P_{dN}\cdot Z) \qquad (1)$$

where $S_{ac}$ is the short circuit capacity, $P_{dN}$ is the rated transmission power, $U_N$ is the rated AC voltage where DC feeds in, $Z$ is the equivalent reactance, normally, line resistance is neglected.

SCR is usually used to measure the connected AC strength from the point view of voltage stability, where HVDC is controlled as constant power and constant extinction angle (CP-CEA) mode. There are two concepts involved in SCR, the critical SCR (CSCR) [4] and the boundary SCR (BSCR) [14].

1) The SCR is called CSCR where the maximum available power (MAP) point coincides with the rated operation point, it can be used to differentiate very weak systems from weak systems.

2) The SCR is called BSCR where the MAP point coincides with the operation point where commutation overlap angle is 30°, it can be used to differentiate weak systems from strong systems. If it is a strong system, the HVDC operation will not be limited by static voltage stability but its own capacity.

Normally, $CSCR \approx 2$ and $BSCR \approx 3$ for the benchmark HVDC model proposed by CIGRE in 1991 (parameters are given in Tab. 1). Then from the practical viewpoint the following conclusions are generally applied: if SCR is less than 2 (CSCR), the AC system is considered as a very weak system. If SCR is greater than 3 (BSCR), the AC system is a strong system. If SCR is between 2 and 3, the AC system is a weak system.

*B. Derivation of Jacobian matrix*

The characteristics of AC/DC voltage stability can be obtained by analyzing Jacobian matrix. In SIDC system, the linearized matrix of AC system can be expressed as

$$\begin{bmatrix} \Delta P/U \\ \Delta Q/U \end{bmatrix} = \begin{bmatrix} J_{p\delta} & J_{pv} \\ J_{q\delta} & J_{qv} \end{bmatrix} \begin{bmatrix} \Delta\delta \\ \Delta U \end{bmatrix} \tag{2}$$

where $J_{p\delta}$, $J_{pv}$, $J_{q\delta}$ and $J_{qv}$ are elements of Jacobian matrix.

Similarly, the linearized matrix of DC system can be expressed as

$$\begin{bmatrix} \Delta P/U \\ \Delta Q/U \end{bmatrix} = \begin{bmatrix} \tilde{J}_{p\delta} & \tilde{J}_{pv} \\ \tilde{J}_{q\delta} & \tilde{J}_{qv} \end{bmatrix} \begin{bmatrix} \Delta\delta \\ \Delta U \end{bmatrix} \tag{3}$$

So after considering the effect of HVDC, the Jacobian matrix of AC/DC system is

$$J = \begin{bmatrix} J_{p\delta} & J_{pv} \\ J_{q\delta} & J_{qv} \end{bmatrix} + \begin{bmatrix} \tilde{J}_{p\delta} & \tilde{J}_{pv} \\ \tilde{J}_{q\delta} & \tilde{J}_{qv} \end{bmatrix} \tag{4}$$

According to the characteristics of LCC-HVDC, there is

$$\tilde{J}_{q\delta} \approx 0 \qquad \tilde{J}_{p\delta} \approx 0 \tag{5}$$

In order to calculate $J_{pv}$ and $J_{qv}$, the equivalent circuit of HVDC is analyzed, given in Fig.1.

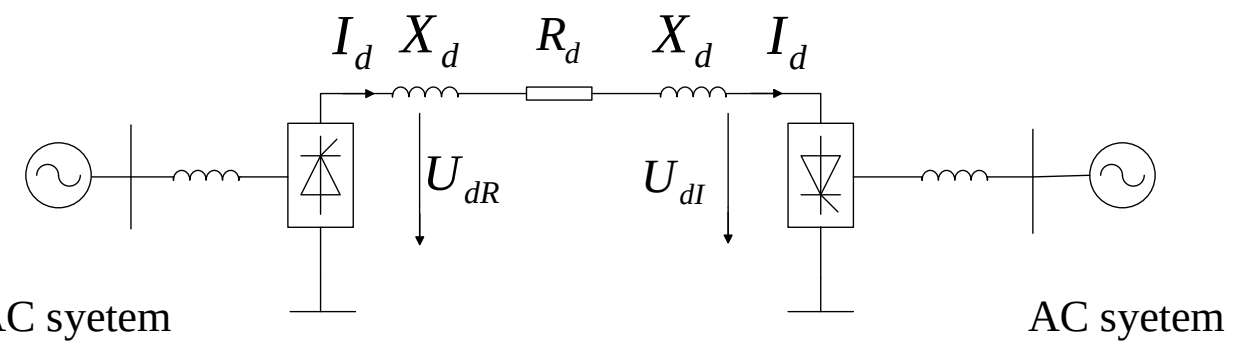

Fig. 1 Equivalent circuit of SIDC system

The equations of inverter side are

$$\begin{cases} P_I = P_d - I_d^2 R \\ U_{dI} = 3\sqrt{2}\pi^{-1} N K_I U_I \cos\gamma - 3\pi^{-1} N X_I I_d \\ \cos\varphi_I = \cos\gamma - \dfrac{X_I I_d}{\sqrt{2} K_I U_I} \end{cases} \tag{6}$$

where $P_I$ is the in-feed HVDC power, $P_d$ is the HVDC transmission power, $R$ is the HVDC line resistance, $U_{dI}$ is the DC voltage, $K_I$ is the ratio of transformer, $U_I$ is the AC voltage, $X_I$ is commutation reactance, $\gamma$ is the extinction angle, $N$ is the number of cascading converter, $\varphi_I$ is the power-factor angle. All variables with subscript $I$ imply that the variables are on inverter side.

Obviously, the reactive power injection of HVDC is

$$Q = -P\tan\varphi + \omega B_c U^2 \tag{7}$$

where $Q$ is the reactive power injection, $B_c$ is the reactive power compensation capacitor.

Substituting $\varphi(U) = \tan\varphi$ into (6) yields:

$$\varphi(U) = \tan\left[\arccos\left(\cos\gamma - \frac{XI_d}{\sqrt{2}KU}\right)\right] \tag{8}$$

Assuming that the HVDC control mode is CP-CEA. On this premise, the sensitivity of reactive power to voltage can be expressed as

$$\frac{\partial\varphi(U)}{\partial U} = K(c)\frac{\partial c}{\partial U} \tag{9}$$

where $K(c) = \dfrac{1}{(\cos\gamma - c)^2}\dfrac{1}{\sqrt{1-(\cos\gamma - c)^2}}$, $c = \dfrac{XI_d}{\sqrt{2}KU}$.

The equation of HVDC power can be derived from (6) as

$$P_d - I_d^2 R = \left(\frac{3\sqrt{2}}{\pi} NKU\cos\gamma - \frac{3}{\pi} NXI_d\right) I_d \tag{10}$$

Simplifying equation (10) by take a derivative of voltage with respect to both side,

$$\begin{aligned} \frac{\partial c}{\partial U} &= \frac{X}{\sqrt{2}KU}\frac{\partial I_d}{\partial U} - \frac{XI_d}{\sqrt{2}KU^2} \\ &= \left( \frac{\frac{1}{2}\cos\gamma}{1 - \frac{\pi R}{3NX} - \frac{1}{2c}\cos\gamma} - c \right)\frac{1}{U} \approx -\frac{2c}{U} \end{aligned} \tag{11}$$

So it can be obtained that

$$\frac{\partial\varphi(U)}{\partial U} = -\frac{2cK(c)}{U} \tag{12}$$

$$\begin{aligned} \tilde{J}_{qv} &= -P(U)\frac{\partial\varphi(U)}{U\partial U} + \frac{2\omega B_c U}{U} \\ &= \frac{P}{U^2}\left(2c\cdot K(c) + \frac{2\omega B_c U^2}{P}\right) = \frac{P}{U^2}T(U,\xi) \end{aligned} \tag{13}$$

where $T(U,\xi) = 2c\cdot K(c) + \dfrac{2\omega B_c U^2}{P}$.

Combining (3) with (10)-(13), the linearized matrix of DC system can be expressed as

$$\begin{bmatrix} \Delta P/U \\ \Delta Q/U \end{bmatrix} = \begin{bmatrix} 0 & 0 \\ 0 & PU^{-2}T(U,\xi) \end{bmatrix} \begin{bmatrix} \Delta\delta \\ \Delta U \end{bmatrix} \tag{14}$$

*C. Relationship between SCR and static voltage stability*

The linearized matrix of AC system can be expressed as (2), where $J_{p\delta} = -EZ^{-1}\cos\delta$, $J_{q\delta} = -P/U$, $J_{pv} = -P/U^2$, $J_{qv} = -Z^{-1}$. Here $E$ is the Thevenin equivalent electric potential and $Z$ is the Thevenin equivalent reactance. Since, at the rated operation point, the reactive power absorbed by HVDC is compensated locally by compensation capacitor. So,

$$U \approx E\cos\delta,\ J_{p\delta} \approx -UZ^{-1} \tag{15}$$

Combining equation (2) with (14), the Jacobian matrix of AC/DC system is

$$J_{sum} := \begin{bmatrix} J_{p\delta} & J_{pv} \\ J_{q\delta} & J_{qv} \end{bmatrix} + \begin{bmatrix} 0 & 0 \\ 0 & \tilde{J}_{qv} \end{bmatrix} \quad (16)$$

It is obvious that, when voltage instability occurs the Jacobian matrix becomes singular.

$$\det\left(J_{sum}\right) = \det\left(\begin{bmatrix} J_{p\delta} & J_{pv} \\ J_{q\delta} & J_{qv} \end{bmatrix} + \begin{bmatrix} 0 & 0 \\ 0 & \tilde{J}_{qv} \end{bmatrix}\right) = 0 . \quad (17)$$

Equation (17) is equivalent to the following equation:

$$\det\left(\tilde{J}_{qv} + J_{qv} - J_{q\delta} J_{p\delta}^{-1} J_{pv}\right) = 0 . \quad (18)$$

Substituting (14) and (15) into (18) and we have

$$\left(\frac{P}{U^2 P_N}\right) T\left(U,\xi\right) + \frac{J_{qv}}{P_N} + \frac{1}{P_N}\frac{P}{U}\frac{Z}{U}\frac{P}{U^2} = 0 . \quad (19)$$

let $\rho = \frac{P}{P_N U^2}$, and as (1) saying that, $SCR = \frac{1}{P_N Z} = -\frac{J_{qv}}{P_N}$, so (19) can be express in another form

$$\Delta(O) = \rho T\left(U,\xi\right) + \rho^2 SCR^{-1} - SCR = 0 . \quad (20)$$

where $O$ represent operation point. It can be seen from (20) that, SCR has a straight relationship with voltage stability.

Considering that $PP_N^{-1} = 1$ and $U = 1$ at the rated operation point, (20) can be simplified as

$$\Delta(O_N) = T\left(U,\xi\right) + SCR^{-1} - SCR = 0 . \quad (21)$$

where $O_N$ represent the rated operation point. As the definition of critical SCR in Section II.A, it satisfies equation (20) when MAP point coincides with the rated operation point.

Similarly, the BSCR is the SCR which satisfies (22) when commutation overlap angle of HVDC is 30°.

$$\Delta(O_B) = T\left(U,\xi\right) + \rho^2 SCR^{-1} - SCR = 0 . \quad (22)$$

where $O_B$ represent the operation point where commutation overlap angle of HVDC is 30°.

It can be concluded from (21) and (22) that both CSCR and BSCR is related to voltage stability margin.

## III. Multi-Infed Generalized Short Circuit Ratio

The original intention of SCR is to measure the strength and stability of AC/DC system. In order to make multi-infeed short circuit ratio having the same clear relevance to voltage stability with SCR, this section continues to use the Jacobian matrix analyzing from the perspective of voltage stability.

All conclusions are derived from the equivalent circuit given in Fig.2 and the following three assumptions are supposed:

**Assumption1**: The DC system is similar, which means that extinction angle and parameters of HVDC based on their individual power capacity have the same values. Besides, the control mode of all HVDC is CP-CEA.

**Assumption2**: The topology of AC system is connected and inductive, namely, the node admittance determinant is reversible and symmetrical.

**Assumption3**: The reactive power absorbed by HVDC is compensated locally, namely, the power on interconnection line is much less than its limitation.

Assumption 1 means that the steady-state characteristics of HVDCs are the same, the only difference is their rated capacity. Since the Jacobian matrix elements of HVDCs are the same based on their rated capacity.

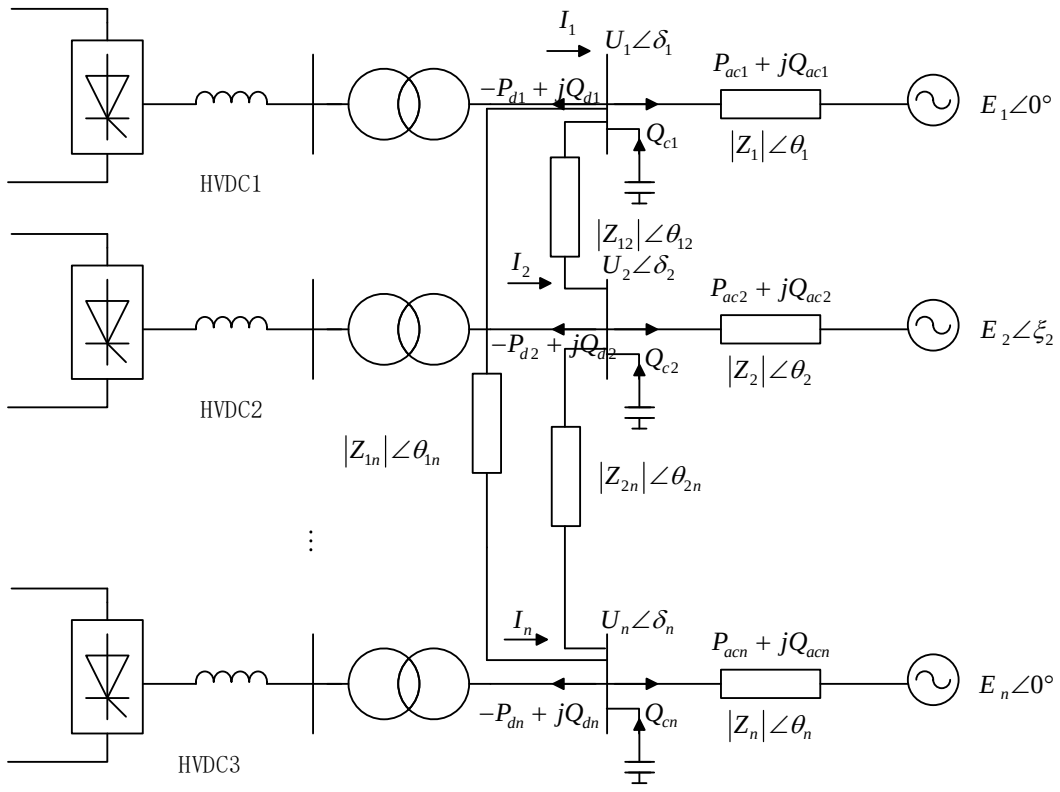

Fig. 2 Typical MIDC system

### A. *Linearization of HVDC system*

As of the SIDC system, the linearization of the *i*th HVDC can be expressed as

$$\begin{bmatrix} \Delta P_i/U_i \\ \Delta Q_i/U_i \end{bmatrix} = \begin{bmatrix} \tilde{J}_{p\delta,i} & \tilde{J}_{pv,i} \\ \tilde{J}_{q\delta,i} & \tilde{J}_{qv,i} \end{bmatrix} \begin{bmatrix} \Delta\delta_i \\ \Delta U_i \end{bmatrix} . \quad (23)$$

where $T_i(\cdot) = 2c_i K\left(c_i\right) + \frac{2}{\rho_i}\frac{\omega B_{ci}}{P_{Ni}}$, $\rho_i = \frac{P_i}{P_{Ni} U_i^2}$, $\tilde{J}_{qv,i} = P_{Ni}\rho_i T_i$, $B_{ci}$ is the compensation capacitor and $P_{Ni}$ is the rated capacity.

The linearization of MIDC system can be written as

$$\begin{bmatrix} \Delta\boldsymbol{P}/\boldsymbol{U} \\ \Delta\boldsymbol{Q}/\boldsymbol{U} \end{bmatrix} \approx \begin{bmatrix} \boldsymbol{0} & \\ & diag\left(P_{Ni}\rho_i T_i\right) \end{bmatrix} \begin{bmatrix} \Delta\boldsymbol{\delta} \\ \Delta\boldsymbol{U} \end{bmatrix} \quad (24)$$

where $\Delta\boldsymbol{U} = [\Delta U_1,...,\Delta U_n]^T$, $\Delta\boldsymbol{Q}/\boldsymbol{U} = [\Delta Q_1/U_1,...,\Delta Q_n/U_n]^T$, diag($a_i$) represents diagonal matrix diag($a_1$,$a_2$,...,$a_n$) for convenience.

### B. *Voltage stability condition*

Similar to (2), the linearization of AC system is

$$\begin{bmatrix} \Delta\boldsymbol{P}/\boldsymbol{U} \\ \Delta\boldsymbol{Q}/\boldsymbol{U} \end{bmatrix} \approx \begin{bmatrix} \boldsymbol{J}_{p\delta} & \boldsymbol{J}_{pv} \\ \boldsymbol{J}_{q\delta} & \boldsymbol{J}_{qv} \end{bmatrix} \begin{bmatrix} \Delta\boldsymbol{\delta} \\ \Delta\boldsymbol{U} \end{bmatrix} . \quad (25)$$

where

$$\begin{bmatrix} \boldsymbol{J}_{p\delta} & \boldsymbol{J}_{pv} \\ \boldsymbol{J}_{q\delta} & \boldsymbol{J}_{qv} \end{bmatrix} = \begin{bmatrix} \boldsymbol{B}diag(U_i) & -diag(P_i/U_i^2) \\ -diag(P_i/U_i) & \boldsymbol{B} \end{bmatrix}$$

here $\boldsymbol{B}$ is the node admittance determinant, $B_{ii} < 0$. So the Jacobian matrix is as follow [15]

$$\boldsymbol{J}_{sys} = \begin{bmatrix} \boldsymbol{B}diag(U_i) & -diag(P_i/U_i^2) \\ -diag(P_i/U_i) & \boldsymbol{B} + diag\left(P_{Ni}\rho_i T_i\right) \end{bmatrix} . \quad (26)$$

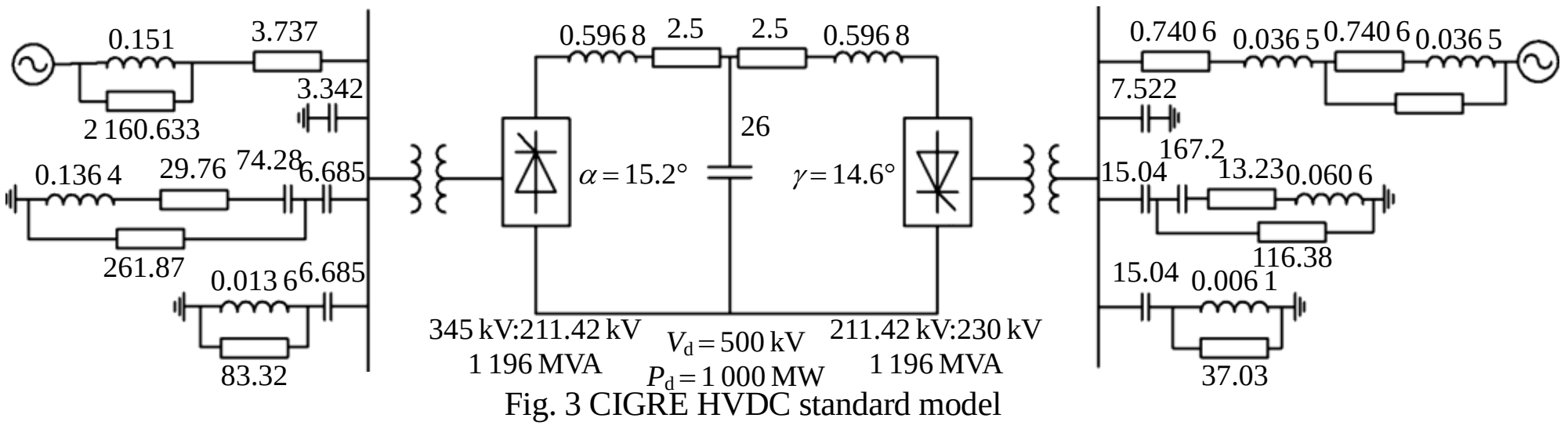

Fig. 3 CIGRE HVDC standard model

Considering that voltage instability implies the singularity of the Jacobian matrix, so by Schur decomposing (26) can be changed to

$$\det\left(\boldsymbol{B}diag(U_i)\right)\det\left(\boldsymbol{J}_{sum}\right)=0\,. \tag{27}$$

where $\boldsymbol{J}_{sum}=\boldsymbol{B}+diag\left(P_{Ni}\rho_i T_i\right)-diag(P_i/U_i^2)B^{-1}diag(P_i/U_i^2)$

$$=\boldsymbol{B}+diag(P_{Ni})diag\left(\rho_i T_i\right)$$

$$-diag(P_{Ni})diag(\rho_i)\left(diag(P_{Ni}^{-1})\boldsymbol{B}\right)^{-1}diag(\rho_i)\,.$$

## C. The generalized short circuit ratio (gSCR)

In order to make multi-infeed short circuit ratio having the same clear physical mechanism with SCR, multiplying $\boldsymbol{J}_{sum}$ left by $diag^{-1}(P_{Ni})$, (27) is equivalent to

$$\det\left[diag\left(\rho_i T_i\right)+diag\left(\rho_i\right)\boldsymbol{J}_{eq}^{-1}diag\left(\rho_i\right)-\boldsymbol{J}_{eq}\right]=0\,. \tag{28}$$

where $\boldsymbol{J}_{eq}$ is defined as the extended Jacobian matrix

$$\boldsymbol{J}_{eq}=-\boldsymbol{DB} \tag{29}$$

Here $\boldsymbol{D}=diag^{-1}(P_{Ni})$.

**Lemma1:** Since the MIDC satisfies **Assumption1** and can be represented by Fig. 1, which means the control mode of HVDCs is CP-CEA, thus the extinction angles of all HVDCs are the same. Normally, the extinction angles are 18° for 50Hz system and 15° for 60Hz system. We can draw a conclusion that, $\rho_i=\rho_j$, $c_i=c_j$ and $B_{ci}P_{Ni}^{-1}=B_{cj}P_{Nj}^{-1}$ is satisfied, so long as the commutation overlap angle (COLA) are the same for all HVDCs, *i.e.* $COLA_i=COLA_j$, while the operation point changes.

**Lemma2:** All Eigenvalues of the matrix $\boldsymbol{J}_{eq}$ are positive. And the minimum eigenvalue of $\boldsymbol{J}_{eq}$ is a simple eigenvalue [16], which means it is unique and its geometric multiplicity and algebraic multiplicity is one, and all elements of its eigenvector are positive too.

Obviously, it can be seen from Lemma1 $\rho=\rho_1=...=\rho_n$, $T\left(U_1,\xi_1\right)=T\left(U_2,\xi_2\right)=...=T\left(U_n,\xi_n\right)$ ,let

$$T\left(U_i,\xi_i\right)\Big|_{i=1,\cdots,n}=2c_iK\left(c_i\right)+\frac{2}{\rho_i}\frac{\omega B_{ci}}{P_{Ni}}=T\left(U,\xi\right) \tag{30}$$

Equation (28) can be simplified as

$$\det\left[\rho T\left(U,\xi\right)\cdot\boldsymbol{I}+\rho^2\boldsymbol{J}_{eq}^{-1}-\boldsymbol{J}_{eq}\right]=0\,. \tag{31}$$

Based on Lemma2, by sorting eigenvalues as $0<\lambda_1<\lambda_2\le\dots\le\lambda_n$, then there exists a nonsingular matrix $\boldsymbol{W}$ such that $\boldsymbol{J}_{eq}$ can be decomposed as follows.

$$\boldsymbol{W}^{-1}\boldsymbol{J}_{eq}\boldsymbol{W}=diag\left\{\lambda_1,\lambda_2,\cdots,\lambda_n\right\}. \tag{32}$$

Thus, it can be deduced from (31) and (32) that:

$$\Pi_{i=1,2,...,n}\left(\rho T\left(U,\xi\right)+\rho^2\lambda_i^{-1}-\lambda_i\right)=0\,. \tag{33}$$

Similarly, equation (33) is equivalent to the following equation from the perspective of voltage stability.

$$\Delta(O)=\rho T\left(U,\xi\right)+\rho^2\lambda_1^{-1}-\lambda_1=0\,. \tag{34}$$

It can be seen that equation (34) is the same as equation (20) mathematically, so equation (34) gives the condition of voltage stability of MIDC system. Therefore $\lambda_1$ can be analogously defined as similar MISCR concept as follows.

**Definition1**: The minimum eigenvalue of the extended Jacobian matrix $\boldsymbol{J}_{eq}$ is defined as the generalized short circuit ratio (gSCR):

$$gSCR=\min\lambda\left(\boldsymbol{J}_{eq}\right). \tag{35}$$

**Definition2**: The gSCR is called the critical gSCR (CgSCR) if the MIDC's rated operation point satisfies (34).

**Definition3**: The gSCR is called the boundary gSCR (BgSCR) if the MIDC's operation point satisfies (34) and its corresponding commutation overlap angle is 30°.

According to Lemma 1, the extinction angle and commutation overlap angle of multiple HVDCs can be kept equivalent respectively by proper control measures, then mathematically the following conclusion can be drawn: $CgSCR\approx CSCR\approx 2$, $BgSCR\approx BSCR\approx 3$. Therefore, based on the definition of gSCR, the strength and stability of AC/DC system can be measured exactly. If gSCR is less than 2, the AC system is a very weak system, which means it can't support all HVDCs operating at the rated operation point. If gSCR is greater than 3, the AC system is a strong system, which can guarantee all the HVDCs to approach to their limit conditions (i.e., the maximum commutation overlap angle is 30°). If gSCR is between 2 and 3, the AC system is a weak system, which means it can support all HVDCs operating at the rated operation point but can't support all HVDCs operating at the point where the commutation overlap angle is 30°.

## IV. Simulation Validation

The system is based on the benchmark HVDC model proposed by CIGRE in 1991, given in Fig.3. Its per unit value is given in Table 1.

Table 1 System per unit value

| | AC on rectifier side | AC on inverter side | DC system |
|---|---|---|---|
| **Voltage** | 345kV | 230kV | 500kV |
| **Power** | 1000MW | 990MW | 1000MW |
| **impedance** | 119.03Ω | 53.43Ω | 250Ω |

In order to validate that $CgSCR\approx 2$ and $BgSCR\approx 3$ are two critical values for strength evaluation in MIDCs, as stated in Definition 2 and 3, in this section, simulations are conducted on both Matlab platform and DIgSILENT platform.

First of all, a dual-infeed system is constructed and simulated on Matlab platform. By changing the rated transmission power of one HVDC, keeping other network parameter un-changed, the system CgSCR and BgSCR are then calculated and plotted in Fig. 4, in which the CgSCR value is about 2 while the BgSCR value is about 3, the same as the values of CSCR and BSCR. Besides, it can be observed that, the deviation of CgSCR and BgSCR is small when the rated power varies. The deviation of CgSCR is about 1.54%, while the deviation of BgSCR is about 0.61%. This means that for a MIDC the characteristics of the voltage stability and the strength of the AC system can be described accurately by the proposed gSCR index.

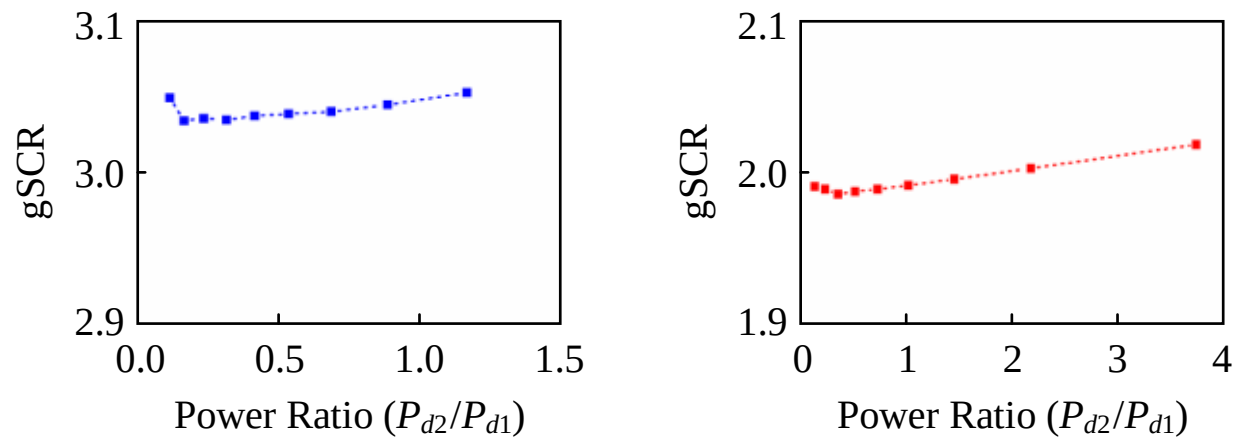

(a) BgSCR (b) CgSCR

Fig. 4 Critical and boundary gSCR of dual-infeed

In order to validate $CgSCR \approx 2$ and $BgSCR \approx 3$ are two critical values for strength evaluation in MIDCs, here we adjust the network equivalent impedance according to our previous work in [17], while keeping other network parameter unchanged, until gSCR=2 and gSCR=3 are satisfied respectively. Observing the characteristics at the voltage instability operation point, namely, confirming the deviation of power between rated and critical point while confirming the deviation of angle between the boundary point and 30°.

Eight simulations have been made and the results are given in Table 2 and Table 3. The voltage instability point is estimated by the divergence condition of power flow on DIgSILENT platform.

Table 2 Critical gSCR and critical power

| Case | Quantity of infeed | gSCR | $Pd_1$/MW | $Pd_2$/MW | $Pd_3$/MW |
|---|---|---|---|---|---|
| **Case1** | Single | 2 | 992.59 | - | - |
| **Case2** | Dual | 2 | 992.89 | 992.22 | - |
| **Case3** | Triple | 2 | 990.23 | 990.19 | 990.19 |
| **Case4** | **$Pd_1$/MW** | **$Pd_1$/MW** | **$Pd_1$/MW** | **$Pd_1$/MW** | **$Pd_1$/MW** |
| | 990.47 | 990.43 | 990.43 | 990.42 | 990.42 |

Table 3 Boundary gSCR of MIDC system

| Quantity of infeed | Case5/Single | Case6/Dual | Case7/Triple | Case8/Triple |
|---|---|---|---|---|
| **gSCR** | 3 | 3 | 3 | 3 |
| **Pd1/MW** | 1037.40 | 1008.72 | 1014.50 | 998.07 |
| **Pd2/MW** | - | 1097.93 | 1052.70 | 1076.48 |
| **Pd3/MW** | - | - | 1052.70 | 1076.48 |
| **OL.A1** | 30.03 | 30.94 | 31.08 | 30.82 |
| **OL.A2** | - | 29.75 | 30.17 | 29.78 |
| **OL.A3** | - | - | 30.17 | 29.78 |

Table 2 shows that when gSCR=2, the maximum available power points of all the HVDCs are around the rated operation points, the deviation of power is less than 0.3%. Furthermore, if gSCR is around 3, the MAPs of all the HVDCs are around the point where the commutation overlap angles are 30°, the deviation of commutation overlap angle is less than 1.1°, which is given in Table 3.

## V. Conclusion

The relationship between the short circuit ratio and static voltage stability is analyzed in this paper for MIDC. The gSCR is proposed to quantitively measure the strength and stability of AC/DC system. Theoretical analysis and simulations show that if gSCR < 2, the AC system is a very weak system; if gSCR > 3, the AC system is a strong system; if gSCR is between 2 and 3, the AC system is a weak system. The gSCR index provides theoretical basis for power system planning and voltage stability analysis. The characteristics of the voltage stability and the strength of the AC system can be described accurately by the proposed gSCR index for a MIDC. Exploring the variety of gSCR under weaker assumptions will be our future work.